\documentclass[aps,pre,twocolumn,showpacs,superscriptaddress]{revtex4}
\usepackage{amsmath}
\usepackage{graphics}
\usepackage[pdftex]{graphicx}
\usepackage{times}
\begin{document}

\title{Solving the collective-risk social dilemma with risky assets in well-mixed and structured populations}

\author{Xiaojie Chen}
\email{xiaojiechen@uestc.edn.cn}
\affiliation{School of Mathematical Sciences, University of Electronic Science and Technology of China, Chengdu 611731, China}

\author{Yanling Zhang}
\affiliation{Center for Systems and Control, College of Engineering, Peking University, Beijing 10087, China}

\author{Ting-Zhu Huang}
\affiliation{School of Mathematical Sciences, University of Electronic Science and Technology of China, Chengdu 611731, China}

\author{Matja{\v z} Perc}
\email{matjaz.perc@uni-mb.si}
\affiliation{Faculty of Natural Sciences and Mathematics, University of Maribor, Koro{\v s}ka cesta 160, SI-2000 Maribor, Slovenia}
\affiliation{Department of Physics, Faculty of Science, King Abdulaziz University, Jeddah, Saudi Arabia}
\affiliation{CAMTP -- Center for Applied Mathematics and Theoretical Physics, University of Maribor, Krekova 2, SI-2000 Maribor, Slovenia}

\begin{abstract}
In the collective-risk social dilemma, players lose their personal
endowments if contributions to the common pool are too small. This
fact alone, however, does not always deter selfish individuals from
defecting. The temptations to free-ride on the prosocial efforts of
others are strong because we are hardwired to maximize our own
fitness regardless of the consequences this might have for the
public good. Here we show that the addition of risky assets to the
personal endowments, both of which are lost if the collective target
is not reached, can contribute to solving the collective-risk social
dilemma. In infinite well-mixed populations risky assets introduce
new stable and unstable mixed steady states, whereby the stable
mixed steady state converges to full cooperation as either the risk
of collective failure or the amount of risky assets increases.
Similarly, in finite well-mixed populations the introduction of
risky assets enforces configurations where cooperative behavior
thrives. In structured populations cooperation is promoted as well,
but the distribution of assets amongst the groups is crucial.
Surprisingly, we find that the completely rational allocation of
assets only to the most successful groups is not optimal, and this
regardless of whether the risk of collective failure is high or low.
Instead, in low-risk situations bounded rational allocation of
assets works best, while in high-risk situations the simplest
uniform distribution of assets among all the groups is optimal.
These results indicate that prosocial behavior depends sensitively
on the potential losses individuals are likely to endure if they
fail to cooperate.
\end{abstract}

\pacs{89.75.Fb, 87.23.Ge, 87.23.Kg}
\maketitle

\section{Introduction}
Cooperative behavior is essential for the maintenance of public
resources and for their preservation for future generations
\cite{ostrom_90, poteete_10, branas2010altruism, grujic_pone10,
capraro2013model, rand_tcs13, levin_pnas14}. However, human
cooperation is often threatened by the lure of short-term advantages
that can be accrued only by means of freeriding and defecting.
Bowing to such temptations leads to an unsustainable use of common
resources, and ultimately such selfish behavior may lead to the
``tragedy of the commons'' \cite{hardin_g_s68}. There exist
empirical and theoretical evidence in favor of the fact that our
climate is subject to exactly such a social dilemma
\cite{schneider_n01, milinski_pnas06,
tavoni_pnas11,milinski_pnas08,milinski_cc11}. And recent research
concerning the climate change has revealed that it is in fact the
risk of a collective failure that acts as perhaps the strongest
motivator for cooperative behavior \cite{milinski_pnas08,
moreira_srep13, vasconcelos_pnas14, vasconcelos_ncc13}.

The most competent theoretical framework for the study of such
problems, inspired by empirical data and the fact that failure to
reach a declared global target can have severe long-term
consequences, is the so-called collective-risk social dilemma
\cite{santos_pnas11, chakra_pcbi12, hilbe_pone13}. As the name
suggests, this evolutionary game captures the fact discovered in the
experiments that a sufficiently high risk of a collective failure
can significantly elevate the chances for coordinating actions and
for altogether avoiding the problem of vanishing public goods.
Recent research concerning collective-risk social dilemmas has
revealed that complex interaction networks, heterogeneity, wealth
inequalities as well as migration can all support the evolution of
cooperation \cite{wang_j_pre09, wang_j_pre10b, du_jm_pre12,
chen_xj_epl12, chen_xj_pre12b, wu_t_pone13, chakra_jtb14} (for a
comprehensive review see \cite{pacheco_plr14}). Moreover,
sanctioning institutions can also promote public cooperation
\cite{sigmund_n10, szolnoki_pre11, perc_srep12, espin2012patient}.
More specifically, it has been shown that a decentralized,
polycentric, bottom-up approach involving multiple institutions
instead of a single, global one provides significantly better
conditions both for cooperation to thrive as well as for the
maintenance of such institutions in collective-risk social dilemmas
\cite{vasconcelos_ncc13}. Voluntary rewards have also been shown to
be effective means to overcome the coordination problem and to
ensure cooperation, even at small risks of collective failure
\cite{sasaki_bl14}. The study of collective-risk social dilemmas can
thus inform relevantly on the mitigation of global challenges, such
as the climate change \cite{inman_ncc09}, but it is also important
to make further steps towards more realistic and sophisticated
models, as outlined in the recent review by Pacheco et
al.~\cite{pacheco_plr14} and several enlightening commentaries that
appeared in response.

Here we consider the collective-risk social dilemma, where in
addition to the standard personal endowments, players own additional
assets that are prone to being lost if the collective target is not
reached. Indeed, individual asset has been considered in the
behavioral experiments regarding the collective-risk social dilemma
\cite{milinski_cc11}. However, different from the experimental study
that investigates the interaction between wealth heterogeneity and
meeting intermediate climate targets \cite{milinski_cc11}, we here
explore in detail whether and how the so-called risky assets provide
additional incentives for individuals to cooperate in well-mixed and
structured populations. It is important to emphasize that, within
our setup, individuals might lose more from a failed collective
action than they can gain if the same action is successful.
Naturally, this constitutes an important feedback for the selection
of the most appropriate strategy. A simple example from real life to
illustrate the relevance of our approach is as follows: Imagine
farmers living around a river that often floods. The farmers needs
to invest into a dam to prevent the floods from causing damage. If
the farmers cooperate and successfully build the dam, they will be
able to enjoy the harvest. However, if the farmers fail to build the
dam, they will lose not only the harvest, but they will also incur
property damage to their fields, houses and stock. Further to the
motivation of our research, it is also often the case that
individuals have limited investment capabilities, which they have to
carefully distribute among many groups. In other words, individuals
may participate in several collective-risk social dilemmas, for
example in each with a constant contribution \cite{santos_n08,
perc_jrsi13}. Rationally, however, individuals tend to allocate
their asset into groups so as to avoid, or at least minimize,
potential losses based on the information concerning risk in the
different groups.

To account for these realistic scenarios, we consider the
collective-risk social dilemma with risky assets in finite and
infinite well-mixed populations, as well as in structured
populations. We first explore how the introduction of risky assets
affects the evolutionary dynamics in well-mixed populations, where
we observe new stable and unstable mixed steady states, whereby the
stable mixed steady state converges to full cooperation in
dependence on the risk. Subsequently, we turn to structured
populations, where the distribution of assets amongst the groups
where players are members becomes crucial. In general, we will show
that the introduction of risky assets can promote the evolution of
cooperation even at low risks, both in well-mixed and in structured
populations, and by doing so thus contributes to the resolution of
collective-risk social dilemmas.

\section{Collective-risk social dilemma with risky assets}

\subsection{Minimal model with risky assets in well-mixed populations}
We first consider the simplest collective-risk social dilemma game
with constant individual assets. From a well-mixed population, $N$
players are chosen randomly to form a group for playing the game. In
the group, each player $y$ with the amount of asset $a$ can choose
to cooperate with strategy $S_y=1$ or defect with strategy $S_y=0$.
Cooperators contribute a cost $c$ to the collective target while
defectors contribute nothing. If all the contributions within the
group either reach or exceed the collective target $T$, each player
$y$ within the group obtains the benefit $b$ $(b>c>0)$, such that
the net payoff is $P_y=b-cS_y$. However, if the collective target is
not reached, all the players within the group lose their investment
and asset with probability $r$, such that the net payoff is then
$P_y=-cS_y-a$, while with probability $1-r$ the payoff remains the
same if the collective target $T$ is reached. Based on these
definitions, the payoff of player $y$ with strategy $S_y$ in a group
with $j$ cooperators is
\begin{eqnarray}
P_y(j)&=&b\theta(j-T)+b(1-r)[1-\theta(j-T)] \nonumber
\\&-&ra[1-\theta(j-T)]-cS_y,
\end{eqnarray}
where $\theta(\omega)=0$ if $\omega<0$ and $\theta(\omega)=1$ otherwise.

We emphasize that an individual will suffer from a risk to lose
everything (the investment and the asset) it has, if the collective
target is not reached in the minimal model. This is in line with the
original definition of the collective-risk social dilemma
\cite{milinski_pnas06,milinski_pnas08,santos_pnas11}. Different from
the original model, however, in our case the asset together with the
investment can be more than the expected benefit of mutual
cooperation. As argued in the Introduction, such scenarios do exist
in reality, and as we will show in the Results section, the risky
assets influence significantly the evolutionary dynamics in both
well-mixed and structured populations. We also refer to the Appendix
for details with regards to the performed analysis.

\subsection{Extended model with asset allocation in structured populations}

\begin{figure*}
\centering
\includegraphics[width=13cm]{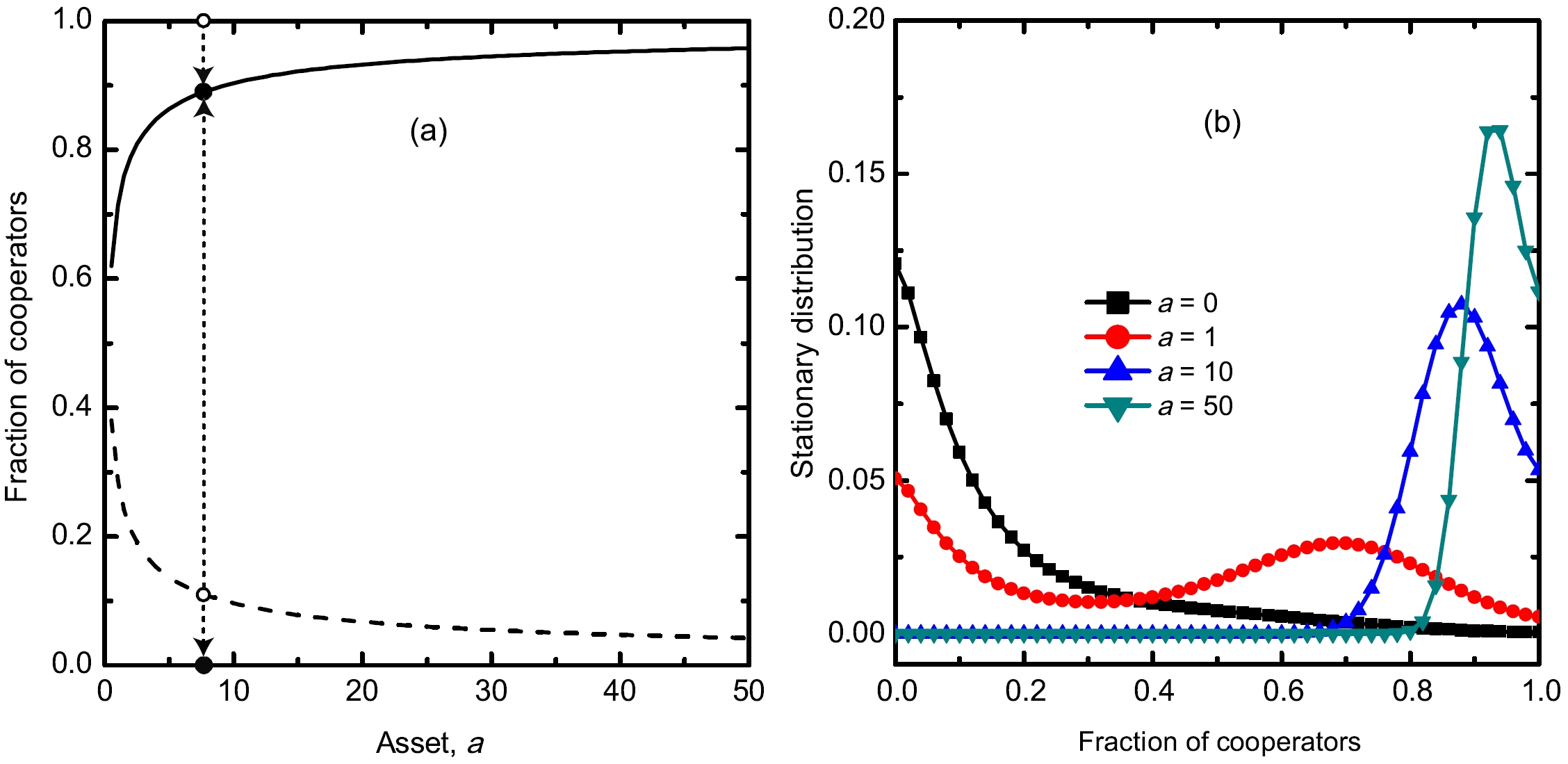}
\caption{(Color online) Evolutionary dynamics of cooperation and
defection in the collective-risk social dilemma with risky assets,
as observed in well-mixed populations. (a) The stable equilibria
(solid line) and unstable equilibria (dash line) as a function of
the asset $a$ in infinite populations. (b) Stationary distribution
of cooperation in finite populations for different values of
individual asset $a$ (see legend) in the presence of mutation $\mu$
(see Appendix for details). Other parameters values are: $N=5$,
$T=3$, $c/b=0.1$, $r=0.2$ in (a), and $N=5$, $T=3$, $Z=50$,
$c/b=0.1$, $r=0.2$, $\mu=0.01$ in (b).} \label{fig1}
\end{figure*}

We here extend the collective-risk social dilemma game with risky
assets to be played on the square lattice with periodic boundary
conditions, where $L^2$ players are arranged into overlapping groups
of size $N=5$ such that everyone is connected to its four nearest
neighbors. Accordingly, each individual belongs to five different
groups and it participates in five collective-risk games. Concerned
for the loss of its assets, each individual $y$ aims to transfer
these assets into those groups that have a lower probability to fail
to reach the collective target. With the information at hand from
the previous round of the game, player $y$ at time $t$ transfers the
asset $a_y^m(t)$ into the group $G_m$ centered on player $m$ according to
\begin{equation}
a_y^m(t)=a\frac{[1-r_m(t-1)]^{\alpha}}{\sum_{n\in
G_y}[1-r_n(t-1)]^{\alpha}},
\end{equation}
where $r_m(t-1)=r$ if at time $t-1$ the number of cooperators
$n_c^m(t-1)<T$ and $r_m(t-1)=0$ otherwise, and $\alpha$ is the
allocation strength of the asset. Here, we mainly consider $\alpha
\geq 0$, given that players generally prefer to allocate their asset
into a relatively safe environment. We note that $\alpha=0$ means
allocating the asset equally into all the groups without taking into
account the information about risk. Accordingly, we will refer to this allocation scheme as uniform or equal. Conversely, $\alpha=+\infty$ means that individuals allocate
their assets only into the most successful groups. We will refer to
this as the fully rational allocation scheme. Lastly, for
$0<\alpha<+\infty$, we have the so-called bounded rational
allocation of assets.

In agreement with the above definitions, the payoff of player $y$ at
time $t$ with strategy $S_y(t)$ and being member of the group that
is centered on player $m$ is
\begin{eqnarray}
P_y^m(t)&=&b\theta[n_c^m(t)-T]+b(1-r)\{1-\theta[n_c^m(t)-T]\} \nonumber
\\&-&ra_y^m(t)\{1-\theta[n_c^m(t)-T]\}-cS_y(t).
\end{eqnarray}
The total payoff at time $t$ is then simply the accumulation of
payoffs from each of the five individual groups where player $y$ is
member, given as $P_y=\sum_{m}P_y^m(t)$.

After the accumulation of payoffs as described above, each player $y$ is allowed to learn from one randomly chosen neighbor $z$ with a probability given by the Fermi function
\cite{szabo_pre98, szabo_pr07}
\begin{equation}
p=[1+\exp^{-\beta(P_z-P_y)}]^{-1},
\end{equation}
where in agreement with the settings in finite well-mixed
populations (see Appendix for details), we use the intensity of
selection $\beta=2.0$. Further with regards to the simulations
details, we note that initially each player is designated either as
a cooperator or defector with equal probability, and it equally
allocates its asset into all the groups in which it is involved.
Monte Carlo simulations are carried out for the population on the
square lattice. We emphasize there exist ample evidence, especially
for games that are governed by group interactions
\cite{szolnoki_pre09c, szolnoki_pre11c, tanimoto_13}, in favor of
the fact that using the square lattice suffices to reveal all the
relevant evolutionary outcomes. Because the system may reach a
stationary state where cooperators and defectors coexist in the
finite structured population in the absence of mutation
\cite{nowak_n92b}, we determine the fraction of cooperators when it
becomes time-independent. The final results were obtained over $100$
independent initial conditions to further increase accuracy, and
their robustness has been tested on populations ranging from
$L^2=2500$ to $10^5$ in size.

\section{Results}

\subsection{Well-mixed populations}

\begin{figure*}
\centering
\includegraphics[width=11cm]{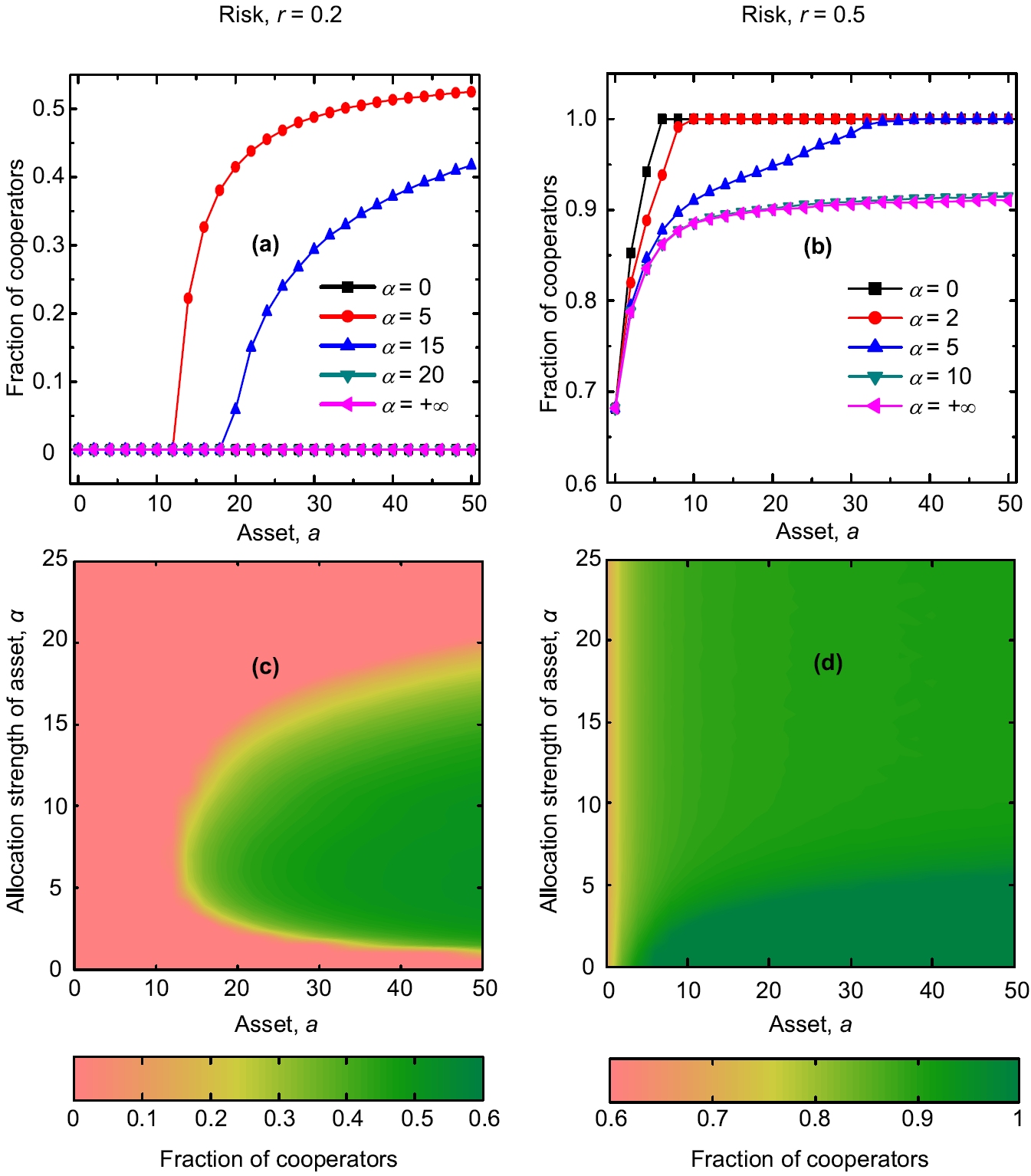}
\caption{(Color online) Evolutionary dynamics of cooperation and
defection in the collective-risk social dilemma with risky assets in
structured populations. Top row depicts the stationary fraction of
cooperators as a function of the asset $a$ for different values of
allocation strength $\alpha$. The risk value is $r=0.2$ in (a) and
$r=0.5$ in (b). Bottom row depicts the contour plot of the fraction
of cooperators as a function of the asset $a$ and the allocation
strength $\alpha$, as obtained for the risk value $r=0.2$ in (c) and
$r=0.5$ in (d). Other parameters values are: $c/b=0.1$, $N=5$, and
$T=3$.} \label{fig2}
\end{figure*}

We begin by showing the results obtained in well-mixed populations,
when individual risky assets are incorporated into the
collective-risk social dilemma as described by the minimal model. In
infinite well-mixed populations, according to the replicator
equation, it can be observed that only the presence of a high risk
leads to two additional, interior equilibria beside the two boundary
equilibria $x=0$ (stable) and $x=1$ (unstable) in the standard
collective-risk social dilemma (Appendix A) \cite{santos_pnas11}.
The unstable interior equilibrium, if it exists, divides the range
$[0, 1]$ of $x$ into two basins of attraction. In other words, when
there is no risk or a low risk, the population has no interior
equilibria, and the only stable equilibrium is at $x=0$,
corresponding to the emergence of full defection
[Fig.~\ref{fig1}(a)]. However, in this unfavorable situation, we
find that the introduction of risky assets can lead to the emergence
of two mixed internal equilibria, which renders cooperation viable
and in fact yields similar effects as a high risk environment.
Furthermore, we find that as the asset increases, the stable
interior equilibrium rapidly increases closely to one, while the
unstable interior equilibrium rapidly decreases closely to zero
[Fig.~\ref{fig1}(a)] (see Appendix A for more analytical results).

In finite well-mixed population, we present the stationary
distribution of cooperation for different values of asset
 $a$ for a population of size $Z=50$, where the group size $N=5$ and the
threshold $T=3$ have been used, as shown in Fig.~\ref{fig1}(b). It
is worth pointing out that the stationary distribution characterizes
the pervasiveness in time of a given configuration of the population
in the presence of behavioral mutations (see Appendix for details).
We see that in the absence of individual assets, the population
spends most of the time in configurations where defectors prevail,
especially at low risks of collective failure. However, when risky
assets are introduced, the population begins to spend more time in
configurations where cooperative behavior thrives. In particular, as
the asset increases to a high value, the population spends most of
the time in configurations where cooperators prevail, while it
spends little time in configurations where defectors prevail.

These results highlight that the introduction of risky assets into
the theoretical model of the collective-risk social dilemma
\cite{santos_pnas11} is found to raise the chances of coordinating
actions and escaping the tragedy of the commons. Accordingly, we
conclude that individual assets significantly enhance the positive
impact of risk on the evolution of cooperation.

\subsection{Structured populations}

\begin{figure*}
\centering
\includegraphics[width=15cm]{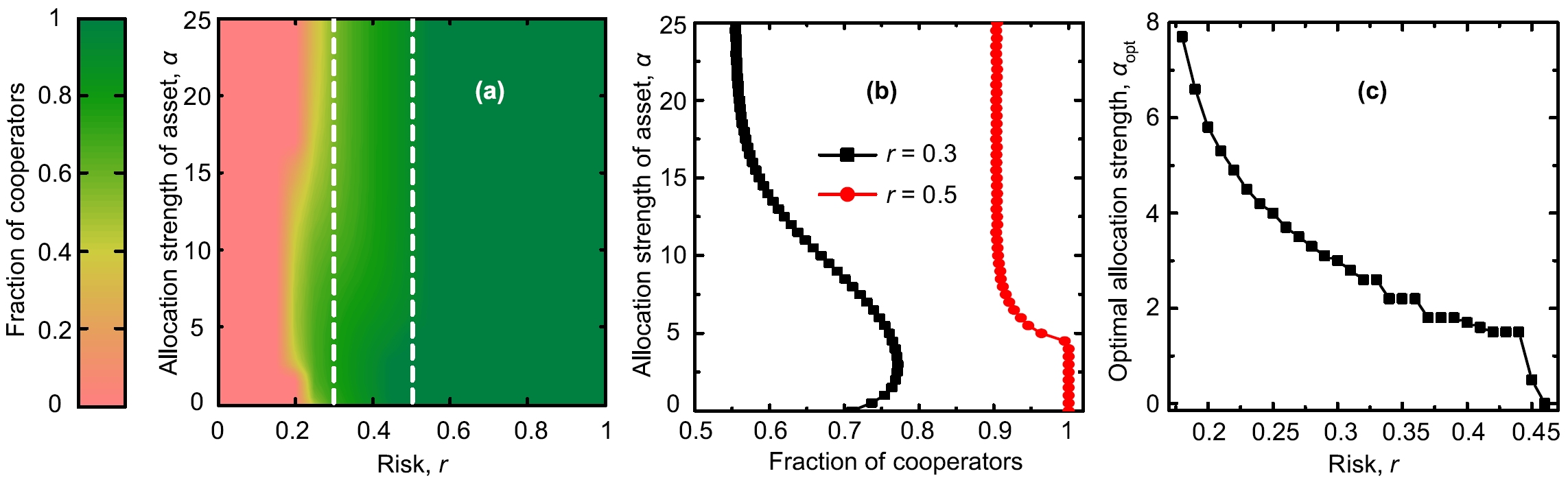}
\caption{(Color online) Successful evolution of cooperation and the
optimal allocation scheme of individual assets. Panel (a) depicts
the stationary fraction of cooperators in dependence on the risk $r$
and the allocation strength $\alpha$, as obtained for $a=25$. Panel
(b) depicts the typical dependence of the fraction of cooperators on
the allocation strength $\alpha$ at $r=0.3$ and $r=0.5$, which are
indicated by white dash lines in panel (a). Panel (c) depicts the
optimal allocation strength $\alpha_{opt}$ that maximizes the
fraction of cooperators in dependence on the risk $r$. Other
parameters values are: $a =25$, $c/b=0.1$, $N=5$, and $T=3$.}
\label{fig3}
\end{figure*}

We continue with presenting the results obtained in structured populations, where at each time step an individual participates in five collective-risk dilemma games and allocates its assets as described by the extended model. Figure~\ref{fig2} shows the fraction of cooperation in the stationary state for two different values of the risk $r$. It can be observed that the fraction of cooperators increases steadily with increasing the value of asset $a$, which is in agreement with the results obtained in well-mixed populations. Moreover, the fraction of cooperators increases with increasing $a$ irrespective of the value of risk $r$ and the allocation strength $\alpha$. More precisely, when the risk is small [Fig.~\ref{fig2}(a) and (c)], defectors always dominate if $a<12$, and this regardless of the value of $\alpha$. On the other hand, if $a>12$, with increasing $\alpha$ the fraction of cooperators first increases, reaches a maximum, but then decreases slowly. This indicates that there exists an optimal allocation strength which can maximize the level of cooperation. In other words, neither uniform allocation nor completely rational allocation is optimal if the risk $r$ is small. Instead, only bounded rational allocation ensures the highest cooperation levels. When the risk is large [Fig.~\ref{fig2}(b) and (d)] the fraction of cooperators still increases with increasing the asset value $a$, but decreases with increasing the allocation strength $\alpha$. Specifically, for $0<a<8$ the fraction of cooperators monotonously
decreases with increasing $\alpha$. For larger assets, however, full cooperation is achieved by uniform allocation, but the same can also be attained with $\alpha$ values that are somewhat larger than zero. But as the value of $\alpha$ increases further, the fraction of cooperators slowly decreases. This can be counteracted by increasing the value of $a$, since then the range of $\alpha$ values where full cooperation is attained increases. Taken together, these results indicate that completely rational allocation of assets is not optimal for the successful evolution of cooperation. Instead, in low-risk situations bounded rational allocation of assets works best, while in high-risk situations the simplest uniform distribution of assets among all the groups is optimal.

In order to study how the reported dependence of the
fraction of cooperators on allocation strength relies on the value
of risk, we first show the stationary fraction of cooperators as a
function of risk and allocation strength together at a certain asset
value $a=25$ in Fig.~\ref{fig3}(a). It can be observed that for risk $r<0.18$
defectors always dominate the population, regardless of the value
of $\alpha$. For $0.18<r<0.46$, there exists an intermediate value
of $\alpha$ that maximizes the fraction of cooperators. For
$0.46<r<0.55$, the fraction of cooperators decreases with increasing
$\alpha$. Finally, for even larger $r$ values, cooperators always
dominate the population. Results presented in fig.~\ref{fig3}(b) show two typical
behaviors depicting the dependence of the cooperation level on
the allocation strength at intermediate values of the risk, as reported
in Fig.~\ref{fig2}. We stress that the range of the risk values for
the typical outcomes depends on other parameters, but the existence
of these results is robust against the variations of the
parameters (see Fig.~\ref{fig5} for details). Moreover, we compute the optimal
value of the allocation strength $\alpha_{opt}$ for several intermediate
risk values, as shown in Fig.~\ref{fig3}(c). We see that
$\alpha_{opt}$ decreases with increasing $r$, and finally reaches
zero. This result highlights that with increasing the risk the
optimal allocation scheme of risky assets gradually translates from bounded rational to the completely uniform allocation.

To explain these results, we continue by showing a series of
snapshots depicting the spatial distribution of strategies over
time. When producing the snapshots we use different colors not just
for cooperators and defectors, but also for distinguishing whether
an individuals' central group is successful or not. More precisely,
blue (yellow) color denotes cooperators (defectors) whose central
group succeeds to reach the collective target. On the other hand,
green (pink) color denotes cooperators (defectors) whose central
group fails to reach the collective target.

\begin{figure*}
\centering
\includegraphics[width=14cm]{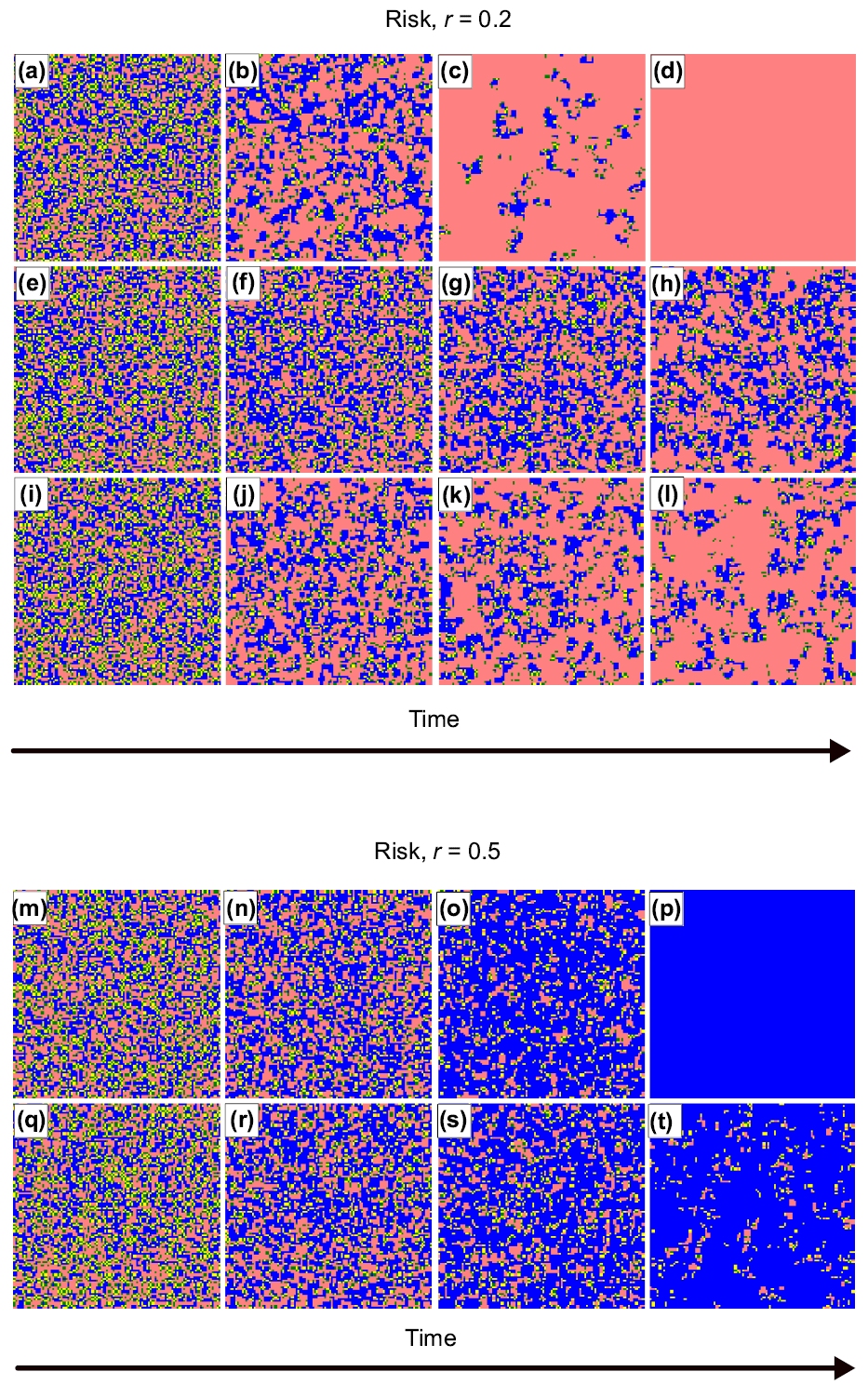}
\caption{(Color online) Spatial pattern distribution as observed
from an randomly initial state at risk $r=0.2$ (top three rows) and
$r=0.5$ (bottom two rows). For $r=0.2$ the allocation strength is
$\alpha=0$ [from (a) to (d)], $\alpha=5$ [from (e) to (h)], and
$\alpha=15$ [from (i) to (l)]. For $r=0.5$ the allocation strength
is $\alpha=0$ [from (m) to (p)] and $\alpha=10$ [from (q) to (t)].
Cooperators whose focal group succeeds (fails) are denoted blue
(green), while defectors whose focal group succeeds (fails) are
denoted yellow (pink). Other parameters are $T=3$, $c/b=0.1$, and
$a=25$.} \label{fig4}
\end{figure*}

\begin{figure*}
\centering
\includegraphics[width=14cm]{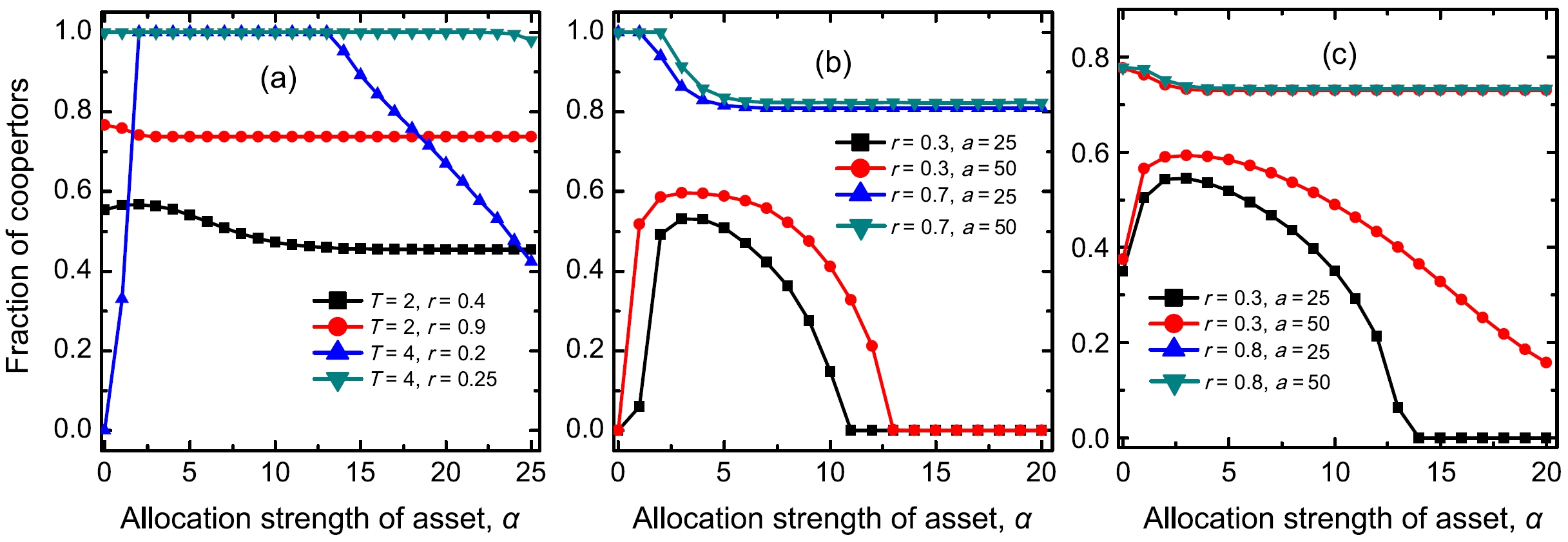}
\caption{(Color online) The robustness of the evolution of
cooperation in the collective-risk social dilemma with risky assets.
Panel (a) depicts the fraction of cooperators as a function of the
allocation strength $\alpha$ for different values of $T$ and $r$, as
obtained for $a=25$, $N=5$, and $c/b=0.1$. Panel (b) depicts the
fraction of cooperators as a function of the allocation strength
$\alpha$ for different values of $r$ and $a$, as obtained for $T=3$,
$N=5$ and a larger $c/b=0.2$ value. Panel (c) depicts the fraction
of cooperators as a function of the allocation strength $\alpha$ for
different values of $r$ and $a$, as obtained on a square lattice
with Moore neighborhood (that is, the group size is $N=9$) for
$a=25$, $c/b=0.1$, and $T=5$.} \label{fig5}
\end{figure*}

In the top three rows of Fig.~\ref{fig4}, we show the representative
sequences for three different values of $\alpha$, all obtained at a
relatively low value of risk. When the risk is small, defectors have
an evolutionary advantage over cooperators \cite{santos_pnas11}, and
they utilize this advantage by gathering the benefit and ultimately
becoming a successful strategy. For low $\alpha$ (top row of
Fig.~\ref{fig4}), both cooperators and defectors allocate the asset
equally to all the groups. The uniform allocation cannot reverse the
direction of invasion of defectors since both cooperators and
defectors may lose a similar amount of asset at a low risk
probability if they put the asset into a predominantly defective
group. Gradually, the success of defectors easily drives the
community into the tragedy of the commons, which is indicated by the
emergence of pink defectors. Note that isolated islands of
cooperators are in the sea of pink defectors, and finally they
disappear completely. For intermediate values of $\alpha$ (second
row of Fig.~\ref{fig4}), individuals tends to put most of their
asset into the successful groups. Thus, islands of cooperators can
more easily preserve their assets than islands of defectors, even if
the risk is low. Hence, cooperators do not lose their assets as much
as defectors. Due to the introduction of risky assets, individual
net payoffs, especially the payoffs from a failed group, depends
strongly on the risk of collective failure. Thus, grouped blue
cooperators are likely to have a higher payoff, at least locally. At
the same time, yellow defectors remain successful if they are in the
vicinity of cooperators, because then they can also preserve most of
their assets and also enjoy most of the benefits from the
surrounding successful groups. But once they invade their
neighboring cooperators, they fast become pink defectors. Although
pink defectors who are around blue cooperators also put most of
their assets into the blue cooperators' central group, and thus
preserve the part of asset, they still put some assets into the
neighboring unsuccessful groups where the centered individual is a
green cooperator or a pink defector. Groups around green cooperators
can partake the assets of pink defectors, which essentially protects
the neighboring blue cooperators. Although the loss of assets
happens with a low probability, it is still sufficiently probable
for cooperators being able to resist the invasion of defectors and
thus to form a mixed dynamical equilibrium in the stationary state.
When $\alpha$ is further increased (third row of Fig.~\ref{fig4}),
defectors will lose a lower amount of their asset at the low risk
probability since they put almost all their assets into the
surrounding successful groups, if only such groups are present. The
increment of $\alpha$ thus sustains the evolutionary advantage of
defectors and the number of blue cooperators consequently decreases.
A few isolated islands of cooperators can withstand the invasion of
defectors because they locally get enough support from the grouped
companions, but if $\alpha$ increases further (not shown), defectors
raise to complete dominance.

The two bottom rows of Fig.~\ref{fig4} show representative snapshot
sequences for two different values of $\alpha$, as obtained at a
relatively high risk. For low values of $\alpha$ (fourth row of
Fig.~\ref{fig4}), blue cooperators do not lose their assets much and
thus manage to have a higher net payoff then defectors. Hence, they
can form compact clusters and expand across the whole population
\cite{chen_xj_epl12}. On the contrary, pink defectors have more
neighbors who are either also pink defectors or green cooperators.
Thus, they lose some of their assets by placing them in the
unsuccessful groups when $\alpha$ is zero or somewhat larger than
zero. Even if $\alpha$ is increased, blue cooperators still lose
less of their asset than other individuals, and can thus still have
the highest net payoff. Therefore they have an evolutionary
advantage, and can eventually dominate the whole population. When
$\alpha$ is sufficiently large (bottom row of Fig.~\ref{fig4}), both
cooperators and defectors tend to allocate their assets
predominantly into the successful surrounding groups. But grouped
defectors still lose some of their assets because they simply do not
have enough neighboring cooperators that would sustain successful
groups. Hence, blue cooperators can still expand across the whole
population, although some tiny specks of defectors manage to survive
by holding on to the blue cooperators. As a result, a few defectors
survive in the sea of cooperators.

To conclude, we report on the robustness of our findings by
investigating changes in dependence on the cost-to-benefit ratio
$c/b$, the collective target $T$, and the interaction structure, as
shown in Fig.~\ref{fig5}. When the threshold value changes, as shown
in Fig.~\ref{fig5}(a), we find that there exists an intermediate
value of the allocation strength $\alpha$ that maximizes the
fraction of cooperators at relatively small values of $r$, while the
fraction of cooperators decreases with increasing allocation
strengths for relatively large values of $r$. Although the range of
$r$ values for the observation of the two typical behaviors varies
as the threshold changes, as indicated by the results presented in
Fig.~\ref{fig5}(a), our main conclusion remain unchanged. Moreover,
in Fig.~\ref{fig5}(b), we find that our results are robust also
against the variations of cost-to-benefit ratio $c/b$. In
particular, as the $c/b$ ratio increases, the evolution of
cooperation is impaired, which is in agreement with previous
research \cite{santos_pnas11,chen_xj_epl12}. But as we increase the
value of the individual assets this trend is again reversed and
cooperative behavior is as pervasive as by low $c/b$ ratios. Lastly,
in Fig.~\ref{fig5}(c), we change the interaction structure by
replacing the Von Neumann neighborhood with the Moore neighborhood.
It can be observed that there still exists an intermediate $\alpha$
value that maximizes the fraction of cooperators at relatively small
values of $r$, and that the fraction of cooperators decrease with
increasing $\alpha$ values at relatively large $r$ values. This
indicates that our main conclusions are robust also against the
changes in the structure of the interaction network.

\section{Discussion}

We have introduced and studied collective-risk social dilemma games
with risky assets, and we have demonstrated how this can lead to
elevated levels of cooperation in well-mixed and structured
populations. The introduction of risky assets increases the stakes
for each individual player, since insufficient contributions to the
common pool result not only in the loss of personal endowments, but
also in the loss of the assets. Thus, players are more prone to
cooperating, and this regardless of their interaction range in the
population. More precisely, we have shown that in infinite
well-mixed populations new stable and unstable mixed steady states
emerge, whereby the stable mixed steady state converges to full
cooperation as either the risk of collective failure or the amount
of risky assets increases. In finite well-mixed populations, we have
shown that the introduction of risky assets drives the population
towards configurations where cooperative behavior abounds. For
comparison, in the absence of risky assets, finite well-mixed
populations are prone to spend the majority of time in
configurations where defectors prevail. In structured populations,
where players have a limited interaction range, we have studied an
extended collective-risk social dilemma games with risky assets,
where the distribution of assets could be tuned by means of an
allocation strength parameter. Among fully rational, bounded
rational and uniform allocation, we have identified the latter as
being optimal for the evolution of cooperation in high-risk
situations. Conversely, in low-risk situations bounded rational
allocation of assets works best. Most surprisingly, the fully
rational allocation of assets only to the most successful groups,
where in principle the assets would be least prone to being lost, is
never optimal. We have explained these results with characteristic
snapshot sequences of strategy distributions in the population, and
we have identified pattern formation as being crucial for the
observed evolutionary outcomes. We have also tested the robustness
of our results with regards to variations of the cost-to-benefit
ratio, the collective target to be reached with the contributions of
players, and with regards to the variation of the interaction
structure, always observing that at least qualitatively our main
conclusions do not change and are fully robust.

As we have emphasized in the Introduction, the consideration
of risky assets in the realm of the collective-risk social dilemma
game is well aligned with reality, in which it is relatively
straightforward to come up with examples where our model could
apply. Our research shows that, at least in theory, such risky or
unsecured, i.e., not immune to loss if the collective target is not
reached, assets significantly promote cooperation and thus
contribute to solving the collective-risk social dilemma. Research based on behavioral experiments has already considered individual assets \cite{milinski_cc11}, although the focus was on the interaction between wealth heterogeneity and intermediate climate targets. The experimental results show that, if players collectively face intermediate climate targets, then rich players are willing to substitute for missing contributions
by poor players \cite{milinski_cc11}. Following this experimental study, a
theoretical work by Abou Chakra and Traulsen \cite{chakra_jtb14}
further showed that rich players contribute on behalf of poor players
only when their own external assets are worth protecting, and moreover, that rich
players maintain cooperation by assisting poor players under a
certain degree of uncertainty. Although the motivation behind our work and the setup are different, we hope that, collectively, the demonstrated importance of individual assets will inspire more research along this line, perhaps in the realm of other evolutionary games or in coevolutionary settings.

\section*{Appendix A: Evolutionary dynamics in infinite well-mixed populations}

For studying the evolutionary dynamics in infinite well-mixed populations, we use the
replicator equation \cite{Hofbauer_98}. To begin, we assume a large
population, a fraction $x$ of which is composed of cooperators, the
remaining fraction $(1-x)$ being defectors. Accordingly, the
replicator equation is
\begin{equation}
\dot{x}=x(1-x)(P_C-P_D),\label{eq1}
\end{equation}
where $P_C$ and $P_D$ are the average payoffs of cooperators and defectors, respectively. Next, let groups of $N$ individuals be sampled randomly from the
population. The average payoff of cooperators is
\begin{equation}
P_C =\sum_{j=0}^{N-1}\left(\begin{array}{c}
N-1\\j\end{array}\right)x^j(1-x)^{N-1-j}P_C(j+1),
\end{equation}
while the average payoff of defectors is
\begin{equation}
P_D =\\ \sum_{j=0}^{N-1}\left (
\begin{array}{c} N-1\\j\end{array} \right)x^j(1-x)^{N-1-j}P_D(j).
\end{equation}

With these definitions, the replicator equation has two
boundary equilibria, namely $x=0$ and $x=1$, whereby full defection is stable while full cooperation is not. Interior equilibria, on the other hand, can be determined by equating $P_C$ and $P_D$, thus obtaining
\begin{equation}
P_C-P_D=\left(\begin{array}{c} N-1\\T-1\end{array}\right)x^{T-1}(1-x)^{N-T}(b+a)r-c=0.
\end{equation}
Furthermore, to determine the interior equilibria, we study the slope and the curvature of the function $G(x)$, which we define as
\begin{equation}
G(x)=\left(\begin{array}{c}
N-1\\T-1\end{array}\right)x^{T-1}(1-x)^{N-T}.
\end{equation}
Note that $P_C-P_D=0$ is equivalent to $G(x)=c/[r(a+b)]$. We thus compute
\begin{equation}
G'(x)=-\left(\begin{array}{c} N-1\\T-1\end{array}\right)x^{T-2}(1-x)^{N-T-1}[1+(N-1)x-T],
\end{equation}
from where it follows that, since $N>2$, $G'(x)$ has a unique internal root at
$\hat{x}=(T-1)/(N-1)$ when $1<T<N$. Moreover, $G'(x)>0$ for
$x<\hat{x}$ and $G'(x)<0$ for $x>\hat{x}$. Accordingly, $G(\hat{x})$ is a unique interior maximum of $G(x)$.

Solving the equation $G(x)=c/[r(a+b)]$ thus yields the
following conclusions:
\begin{enumerate}
\item When $G(\hat{x})>c/[r(a+b)]$,
Eq.~(\ref{eq1}) has two interior equilibria, denoted by $x_{1}^{*}$
and $x_{2}^{*}$ with $x_{1}^{*}<\hat{x}<x_{2}^{*}$. Since $G'(x)>0$
for $x<\hat{x}$ and $G'(x)<0$ for $x>\hat{x}$, $x_{1}^{*}$ is an
unstable equilibrium
and $x_{2}^{*}$ is a stable equilibrium.
\item When $G(\hat{x})=c/[r(a+b)]$, Eq.~(\ref{eq1}) has only one
interior equilibrium $\hat{x}$, which is a tangent point, and is thus
unstable.
\item When $G(\hat{x})<c/[r(a+b)]$, Eq.~(\ref{eq1}) has no interior
equilibria.
\end{enumerate}

When $T=1$ or $T=N$, however, for $c/[r(a+b)]\geq 1$ Eq.~(\ref{eq1}) has no
interior equilibria. While for $c/[r(a+b)]<1$, Eq.~(\ref{eq1}) has
only one interior equilibrium $x^{*}$. Note that
$x^{*}=1-\{c/[r(a+b)]\}^{1/(N-1)}$ is stable for $T=1$ since
$G'(x^{*})<0$, while $x^{*}=\{c/[r(a+b)]\}^{1/(N-1)}$ is unstable for
$T=N$ since $G'(x^{*})>0$.

\section*{Appendix B: Evolutionary dynamics in finite well-mixed populations}

For studying the evolutionary dynamics in finite well-mixed populations, we consider a population of finite size $Z$. Here the average payoffs of cooperators and defectors in the population
with $k$ cooperators are respectively given by
\begin{eqnarray}
f_C(k)&=&\left (
\begin{array}{c} Z-1\\N-1\end{array} \right)^{-1}\sum_{j=0}^{N-1} \left (
\begin{array}{c} k-1\\j\end{array} \right) \left (
\begin{array}{c} Z-k\\N-j-1\end{array} \right)\nonumber
\\&\cdot&P_C(j+1),
\end{eqnarray}
and
\begin{eqnarray}
f_D(k)&=&\left (
\begin{array}{c} Z-1\\N-1\end{array} \right)^{-1}\sum_{j=0}^{N-1} \left (
\begin{array}{c} k\\j\end{array} \right) \left (
\begin{array}{c} Z-k-1\\N-j-1\end{array} \right)\nonumber
\\&\cdot&P_D(j).
\end{eqnarray}
Next, we adopt the pair-wise comparison rule to study the
evolutionary dynamics, based on which we assume that player $y$
adopts the strategy of player $z$
 with a probability given by the
Fermi function
\begin{equation}
p=[1+e^{-\beta(P_z-P_y)}]^{-1},
\end{equation}
where $\beta$ is the intensity of selection that determines the level of uncertainty in the strategy adoption process \cite{szabo_pre98, szabo_pr07}. Without loosing generality, we use $\beta=2.0$ throughout this work.

With these definitions, the probability that the number of cooperators in the
population increases or decreases by one is
\begin{equation}
T^{\pm}(k)=\frac{k}{Z}\frac{Z-k}{Z}[1+e^{\mp\beta[f_C(k)-f_D(k)]}]^{-1}.
\end{equation}
Following previous research \cite{santos_pnas11}, we further introduce the
mutation-selection process into the update rule, and compute the
stationary distribution as the key quantity that determines the
evolutionary dynamics in finite well-mixed populations. We note that, in the presence of
mutations, the population will never fixate in any of the two
possible absorbing states. Thus, the transition matrix of the
complete Markov chain is
\begin{equation}
M=[p_{u,v}]_{(Z+1)\times(Z+1)}^{T},
\end{equation}
where $p_{u,v}=0$ if $|u-v|>1$, otherwise
$p_{u,u+1}=(1-\mu)T^{+}(u)+\mu(Z-u)/Z$,
$p_{u,u-1}=(1-\mu)T^{-}(u)+\mu u/Z$ and
$p_{u,u}=1-p_{u,u+1}-p_{u,u-1}$. Accordingly, the stationary
distribution of the population, that is the average fraction of the
time the population spends in each of the $Z+1$ states, is given by
the eigenvector of the eigenvalue $1$ of the transition matrix $M$
\cite{karlin_75}. In the Results Section, Fig.~\ref{fig1}(b) is
obtained by using $\mu=0.01$.

\begin{acknowledgments}
This research was supported by the National Natural Science
Foundation of China (Grant No. $61203374$ and No. $61370147$), the
National 973 Program of China (Grant No. 2013CB329404), and the
Slovenian Research Agency (Grants J1-4055 and P5-0027). M.P. also acknowledges funding by the Deanship of Scientific Research (DSR), King Abdulaziz University, under grant No. (76-130-35-HiCi).
\end{acknowledgments}

\end{document}